\documentclass[aps,jcp,reprint,groupedaddress, onecolumn]{revtex4-2}

\usepackage{graphicx}
\graphicspath {{Figures/}}

\usepackage{placeins} 
\usepackage{caption}
\usepackage{subcaption}  

\usepackage{bm}
\usepackage{amssymb} 
\usepackage{amsmath}

\usepackage{color}
\usepackage{setspace}   
\usepackage{comment}
\usepackage{soul} 
\usepackage{vector}  

\begin{document}

	
	\title{Bulk viscosity of dilute monatomic gases}
	\author{Bhanuday Sharma}

	\affiliation{Indian Institute of Technology Kanpur, India}
	

	\author{Rakesh Kumar}
	\email[]{rkm@iitk.ac.in}
	\homepage[]{www.iitk.ac.in/~rkm}
	\affiliation{Indian Institute of Technology Kanpur, India}
	
	\author{Savitha Pareek}
	\affiliation{Dell HPC and AI Innovation Lab}
	
	\author{Ashish Singh}
	\affiliation{Dell HPC and AI Innovation Lab}
	
	
	\begin{abstract}
		Extensive research has been carried out in the past to estimate the bulk viscosity of dense monatomic fluids; however, little attention has been paid to estimate the same in the dilute gas regime. In this work, we perform precise Green-Kubo calculations in molecular dynamics simulations to estimate the bulk viscosity of dilute argon gas. The investigated temperature and pressure range is 300 to 750 K and 0.5 to 1.5 bar respectively. It is observed that the estimated bulk viscosity is $O(10^{-10})$ Pa~s, which is several orders of magnitude smaller than that of other diatomic and polyatomic gases, but nonetheless, not an absolute zero as typically assumed. It implies that Stokes' hypothesis is true for dilute monatomic gases only in an approximate rather than absolute sense. The variation of bulk viscosity with pressures at constant temperatures has also been studied and is found to be of quadratic nature. The obtained bulk viscosity values are also reported in the reduced Lennard-Jones units to enable extension of the present results to other noble gases as well. It has also been observed that the bulk viscosity of Lennard-Jones monatomic gases becomes temperature independent at very low densities.
	\end{abstract}
	
	
	\maketitle
	
\section{Introduction}
For an isotropic, homogeneous, and Newtownian fluid, the mechanical, $P_{\text{mech}}$, and thermodynamic pressures, $P_{\text{thermo}}$, are related to bulk viscosity, $\mu_b$, and velocity of fluid, $U$, as follows \cite{gad1995questions}- 
\begin{equation}
	P_{\text{mech}}= P_{\text{thermo}}-\mu_b \nabla\cdot\vec{U} 
	\label{eq:1}
\end{equation}
The mechanical or hydrodynamic pressure, $P_{\text{mech}}$, is the actual pressure caused by the combined effect of random thermal motion of molecules and intermolecular forces. Any pressure measuring instrument placed in the flow-field measures this quantity. The thermodynamic or hydrostatic pressure, $P_{\text{thermo}}$, is hypothetical pressure that would exist if the fluid element is brought to equilibrium through an adiabatic process \cite{okumura2002new}. 
In his original work, Stokes {\em et al.}\cite{stokes1880theories} simplified Eq.~\eqref{eq:1} by assuming the bulk viscosity to be zero. 
Since then, it has become customary to use this approximation, popularly named as the Stokes' hypothesis, in fluid dynamical analysis. Stokes backed his approximation by arguing that, though in the case of uniform dilatation, there might exist an additional normal pressure over and above the equilibrium pressure, $P_{\text{thermo}}$, this additional pressure would be neutralized as the molecular displacements causing this additional pressure do not have any directional preference. However, he also acknowledged that this assumption might be incorrect; and even if theory and experiments agree, it could be due to the overall small value of this additional pressure rather than bulk viscosity being zero. The Navier-Stokes' equation derived with this assumption of zero bulk viscosity failed to accurately predict the sound absorption coefficient for all fluids except dilute monatomic gases \cite{prangsma1973ultrasonic, karim1952second, winter1967high-temperature-ultasonic-measurements}. The experimental investigations on the absorption of sound waves showed an excess absorption over and above that due to shear viscosity, heat conductivity, and radiation. Hence, it was argued that Stokes' hypothesis is not true in general except for the case of dilute monatomic gases. Tisza\cite{tisza1942supersonic} suggested that this excess absorption can be explained if Stokes' hypothesis is ignored in Navier-Stokes' equation, and it is assumed that the effects of all the dissipative mechanisms other than shear viscosity, heat conductivity and radiation can be combined in one parameter \cite{karim1952second, truesdell1954present}. With these assumptions, one can choose a suitable value of bulk viscosity parameter such that the theoretical absorption coefficient matches with the experimental one \cite{prangsma1973ultrasonic}. Later, this approach became the de-facto choice for calculating bulk viscosity coefficient. 

In the past, extensive research \cite{hoover1980bulk, heyes1984thermal, hoheisel1987Bulk-viscosity-of-the-Lennard-Jones-fluid, bertolini2001generalized, fernandez2004molecular, meier2005transport, baidakov2014metastable, jaeger2018bulk} has been carried out to understand and estimate bulk viscosity of dense monatomic fluids; however, a little attention has been paid to bulk viscosity of the same in the dilute gas regime.
Nettleton\cite{nettleton1958intrinsic} used Born-Green theory with Kirkwood superposition approximation to investigate bulk viscosity of dilute monatomic gas and showed that even dilute gases may have very small but nonzero bulk viscosity. He predicted the bulk viscosity of dilute argon gas to be of the order of 1 micropoise or 10$^{-7}$ Pa~s. 

Meier {\em et al.}\cite{meier2005transport} used Green-Kubo formulation to calculate bulk viscosity of Lennard-Jones (LJ) fluid for a wide range of conditions, given in terms of reduced densities ($\rho^*$) and reduced temperatures ($T^*$). These reduced properties are defined in reduced LJ unit system, which is described in Section~\ref{sec:simulation details}. They characterized the pressure autocorrelation function (ACF) and its integral of autocorrelation function (IACF). They observed an oscillatory nature in ACF and provided an explanation on the basis of formation of bound states. 
However, in the gaseous regime, their reported values were limited to reduced temperature 1.2, which in the case of argon gas corresponds to 140~K.
Recently, Jaeger {\em et al.}\cite{jaeger2018bulk} has also reported bulk viscosity of Lennard-Jones fluid for reduced densities greater than 0.1 at reduced temperatures of 1.35, 1.7, and 2.5.
At $\rho^* = 0.1$ and $T^*=2.5$, argon belongs to 300 K and 104.7 bar. At these conditions, argon remains in a supercritical phase.
%
To the best of our knowledge, no attempt has been made so far for the exact quantification of bulk viscosity of dilute monatomic gases at atmospheric conditions; or dilute LJ fluid at reduced densities less than 0.001. 

In this communication, we use Green Kubo formulation to calculate precise values of bulk viscosity of dilute argon gas in the pressure and temperature range of 0.5 to 1.5 bar and 300 to 750 K. Through these results, we highlight the fact that even dilute monatomic gases have small but nonzero bulk viscosity. Hence, Stokes' hypothesis is true for these fluids in an approximate rather than absolute sense, even in the regime where they can be safely considered dilute.
\section{Simulation details}\label{sec:simulation details}

We have performed molecular dynamics (MD) simulations \cite{rapaport2004art, allen2017computer} and used the Green-Kubo approach to estimate the bulk viscosity of dilute argon gas.
Argon atoms are modeled with the standard Lennard-Jones potential $V(r)$-
\[
V(r) =
\begin{cases}
	4 \epsilon \left[ {\left(  {\frac{\sigma}{r}} \right)}^{12} - {\left( \frac{\sigma}{r }\right) }^6   \right],  &  \text{if} ~~r<r_{ \text{cut off}}\\
	0 & \text{if}~~ r>r_{\text{cut off}}\\
\end{cases}
\]
where $r$ is the distance between two atoms, $\sigma$ and $\epsilon$ are Lennard-Jones length and energy parameters respectively, and $r_{\text{cutoff}} = 12$ \AA. The table~\ref{table:Ar-pot-list} gives the details of molecular models used to represent argon atom in this work.	Some of the results are reported here in reduced LJ unit system. These quantities are marked with a superscript asterisk $^*$ and calculated as $T^* = \frac{T}{(\epsilon/k_B)}$, $\rho^* = \rho \frac{\sigma^3}{m}$, $\mu^* = \mu \frac{\sigma^2}{\sqrt{\epsilon m}}$, and $\mu_b^* = \mu_b \frac{\sigma^2}{\sqrt{\epsilon m}}$. Here, $T$, $\rho$, $m$, and $\mu$ are temperature, density, mass of atom, and shear viscosity, respectively. 

The Green-Kubo relations \cite{green1954markoff, kubo1957statistical} relate viscosity coefficients to integral of autocorrelation of pressure tensor.
The shear viscosity, $\mu$, and bulk viscosity, $\mu_b$, are given as $\mu = \lim_{t \to \infty } \mu(t)$, and $\mu_b = \lim_{t \to \infty } \mu_b(t)$, where $\mu(t)$ and $\mu_b(t)$ are given as follows- 
\begin{equation}
	\mu(t) = \frac{\mathcal{V}}{k_B T} \int_{0}^{t} \langle P_{i j}(t')~P_{i j}(0) \rangle dt'  	\label{eq:Green-Kubo_shear-viscosity} 
\end{equation}
\begin{equation}
	\mu_b(t) = \frac{\mathcal{V}}{k_B T} \int_{0}^{t} \langle \delta P(t')~\delta P(0) \rangle dt' \label{eq:Green-Kubo_bulk-viscosity}
\end{equation}

Here, $\mathcal{V}$ is volume of simulation domain, $T$ is the temperature, $P_{ij}$ is $ij^{\text{th}}$ component of pressure tensor at a time $t'$. The angle brackets $\langle \rangle$ represent the ensemble average, i.e., statistical average over many time windows. $\delta P(t')$ is the deviation of instantaneous pressure from mean pressure and is given as $\delta P(t') = P(t') -P_{eq} $. Here, $P(t')$ is the instantaneous value of average of diagonal terms of pressure tensor at time $t'$, i.e., $P(t')=\frac{1}{3}[P_{ii}(t')+P_{jj}(t')+ P_{kk}(t')]$ and $P_{eq}$ is equilibrium pressure of system calculated by taking time average of $P(t')$ over long times. $P(t')$ and $P_{eq}$ essentially represent mechanical and thermodynamic pressures, respectively. 
To obtain the precise value of viscosity, we have used the time decomposition scheme for the calculation of ensemble average of autocorrelation function as suggested by Zhang {\em et al.}\cite{zhang2015reliable-time-decomposition-method}. We have generated 120 trajectories, each with 1 ns and 3 ns of equilibration and production run lengths, respectively. Each of these trajectories contains 200 molecules. Final results are generated by taking the average over all these 120 trajectories. All the simulations are carried out using the open-source software, LAMMPS (Large-scale Atomic/Molecular Massively Parallel Simulator) \cite{plimpton2007lammps}, while all the plots are generated using the open-source software VEUSZ \cite{sanders2008veusz}.

\begin{table}
	\centering
	\begin{tabular}{|c| c| c|}
		\hline
		\rule{0pt}{1.3 em} \textbf{Source} & $\bm{\sigma}$ & $\bm{\epsilon}$ \\[0.2 em]		\hline
		\rule{0pt}{1.3 em} Ref.~\cite{fincham1983comparisons} & 3.405 & 119.8  \\	[0.2 em]	\hline
		\rule{0pt}{1.3 em} Ref.~\cite{rowley1997diffusion} & 3.418 & 124 \\	[0.2 em]	\hline
		\rule{0pt}{1.3 em} Ref.~\cite{kirova2015viscosity} & 3.4 & 120 \\ [0.2 em]		\hline
		\rule{0pt}{1.3 em} Ref.~\cite{fernandez2004molecular} & 3.3952 & 116.79 \\	[0.2 em]	\hline
	\end{tabular}
	
	\caption{Argon atom models evaluated in the present study.}
	\label{table:Ar-pot-list}
\end{table}

\begin{table}
	\centering
	
	\begin{tabular}{|c| c| }
		\hline
		\rule{0pt}{1.3 em} $\bm{T^*}$ & $\bm{\mu^*}$ \\ \hline
		\rule{0pt}{1.3 em} 2.5 & 0.24745 \\ \hline
		\rule{0pt}{1.3 em} 4.2 & 0.38619 \\ \hline
		\rule{0pt}{1.3 em} 5   & 0.4201 \\ \hline
		\rule{0pt}{1.3 em} 6   & 0.48241 \\ \hline
	\end{tabular}
	\caption{Shear viscosity of dilute LJ fluid at $\rho^* = 0.0015$}
	\label{table:shear-viscosity-of-LJ-fluid}
\end{table}

\section{Results and discussion}


\subsection{Selection of atomic potential for dilute argon gas}
There are several atomic potentials available in the literature for modeling argon atom. We have considered four Lennard-Jones potentials (see table~\ref{table:Ar-pot-list}) and evaluated their ability to reproduce the shear viscosity in order to identify the one which represents the gas phase of argon best. 
In this regard, we performed Green-Kubo molecular dynamics simulations at four reduced temperatures, viz., 2.5, 4.2, 5, and 6. The obtained shear viscosity values are given in table~\ref{table:shear-viscosity-of-LJ-fluid}. These reduced temperatures were chosen such that corresponding physical temperatures are close to 300, 400, 600, 700 K, although the exact values of physical temperatures depend on the $\epsilon$ of the potential. The obtained results are used to calculate the shear viscosity of dilute argon gas using the formulae given in Section~\ref{sec:simulation details}. 

\begin{figure}
	\centering
	\includegraphics[width=0.5\linewidth]{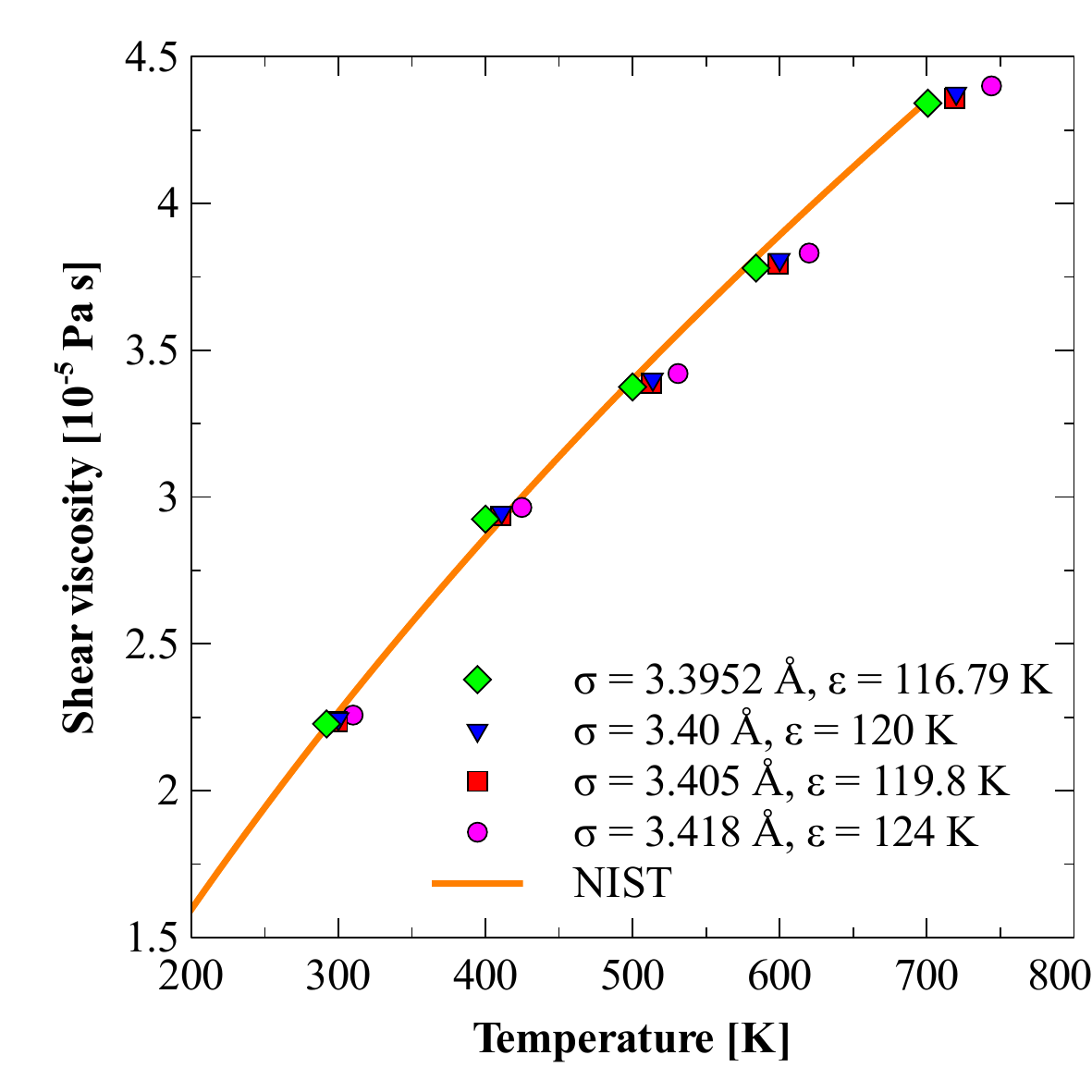}
	\caption{Performance of various molecular dynamics potentials in prediction of shear viscosity of dilute Ar gas. Solid line indicates the reference data obtained from the NIST chemistry WebBook~\cite{Lemmon2021nist}.
	}
	\label{fig:shearviscosityofarvariouspotcombined}
\end{figure}

Figure~\ref{fig:shearviscosityofarvariouspotcombined} shows shear viscosities calculated using these four potentials and compares them with NIST dataset \cite{Lemmon2021nist}. It can be observed that the potential with $\sigma = 3.3952$ \AA \ and $\epsilon/k_B = 116.79$ K \cite{vrabec2001set, fernandez2004molecular}
gives the shear viscosity closest to NIST dataset, and hence, represents the gas phase of argon the best. The same is used in the rest of the simulations performed in this work.



\subsection{Bulk viscosity of dilute argon gas}
\begin{figure}
	\centering
	\begin{subfigure}{0.48\textwidth}
		\centering
		\includegraphics[width=0.9\linewidth]{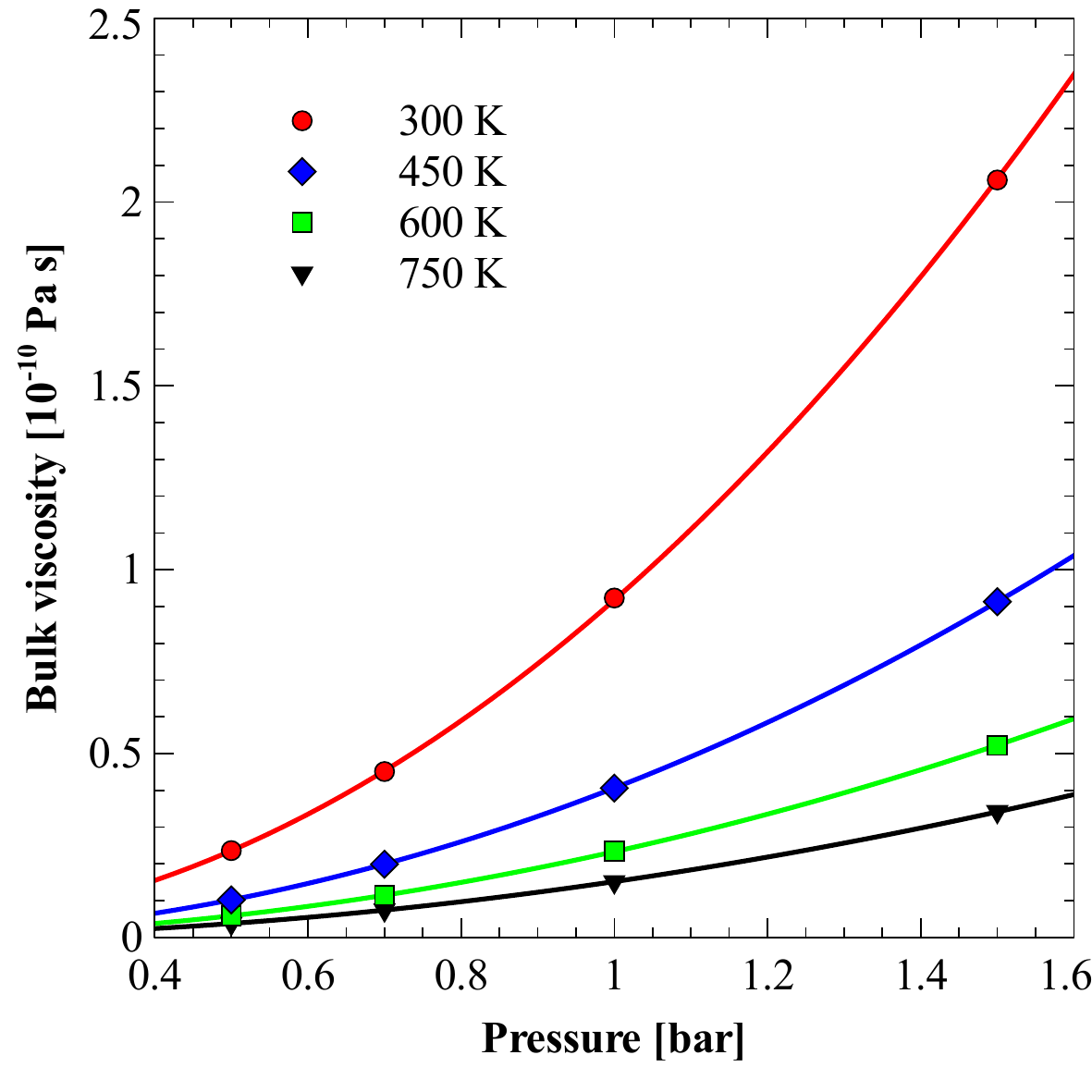}
		\caption{}
		\label{fig:variationwithpressureatalltemp}
	\end{subfigure}
	\begin{subfigure}{0.48\textwidth}
		\centering
		\includegraphics[width=0.9\linewidth]{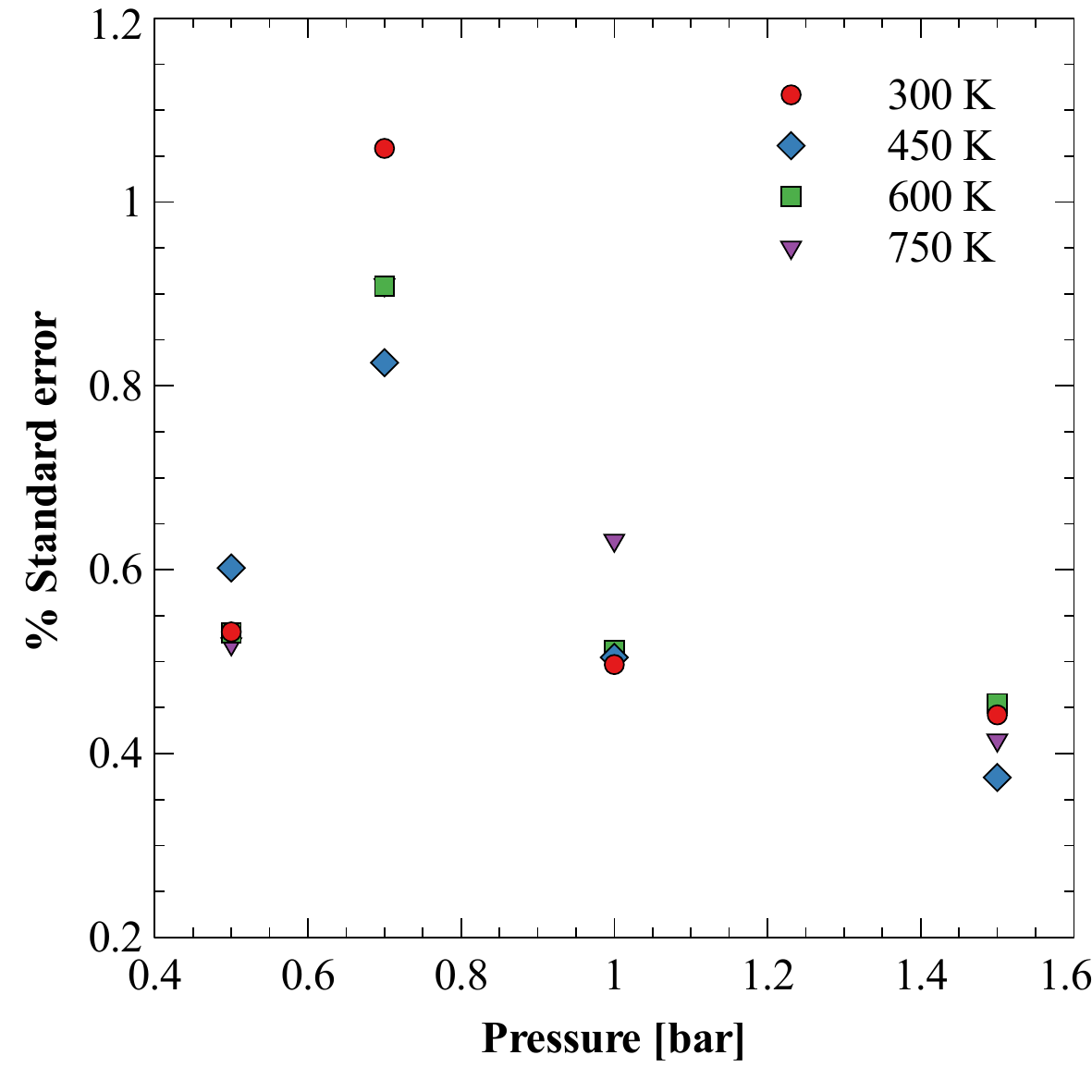}
		\caption{}
		\label{fig:standard-error-in-bulk-viscosity-of-argon}
		
	\end{subfigure}
	\caption{(a) Variation of bulk viscosity of dilute argon gas with pressure at various temperatures.	The data-points are fitted with quadratic curve of shape $y = ax^2+bx+c$. The details of curve fitting, viz., value of $a$, $b$, $c$, and goodness of fit $\chi^2$ are given in table~\ref{table:fitting parameters}. (b)~Standard error in the bulk viscosity values reported in figure~\ref{fig:variationwithpressureatalltemp}.}
\label{fig:bulkviscosityofargon}
\end{figure}

To calculate the bulk viscosity of dilute argon gas, we have considered four different pressures, i.e., 0.5, 0.7, 1.0, and 1.5 bar. For each pressure, simulations were performed at four different temperatures, i.e., 300, 450, 600, and 750 K. The argon gas can be considered dilute in this pressure and temperature range. 
The obtained variation of bulk viscosity with pressure and temperature are plotted in figure~\ref{fig:variationwithpressureatalltemp}. This figure shows that the dilute argon gas has nonzero bulk viscosity values at the all simulated temperatures and pressures. 
It also shows that bulk viscosity decreases as either the temperature increases or the pressure decreases. 
This is because the gas becomes more and more dilute with the increasing temperature and decreasing pressure, thereby making dense gas effects less important. 
This trend is different from that for other gases, such as dilute nitrogen gas, where a weak pressure dependency and linear temperature dependency are observed at similar pressures and temperatures \cite{sharma2019estimation, cramer2012numerical}. The reason for this contrasting behavior is that both gases have different source mechanisms of bulk viscosity, as discussed in the next paragraph.  
It should also be noted that the obtained bulk viscosity values for argon gas are $O(10^{-10})$, which is five order of magnitude smaller than bulk viscosity of most often used diatomic gases, such as dilute nitrogen gas, which has a bulk viscosity of the order of $O(10^{-5})$ \cite{cramer2012numerical}. Since each data point in figure~\ref{fig:variationwithpressureatalltemp} is obtained by taking the average over independent 120 trajectories, the standard error can be calculated to assess the precision of reported values, and the same is plotted in figure~\ref{fig:standard-error-in-bulk-viscosity-of-argon}. It can be seen in this figure that the standard errors are less than 1.1\% for all reported values. The small value of error signifies that the reported bulk viscosity values actually represent a signal (even though small) rather than a manifestation of noise.   

\begin{figure*}
\centering
\begin{subfigure}{0.48\textwidth}
	\centering
	\includegraphics[width=0.9\linewidth]{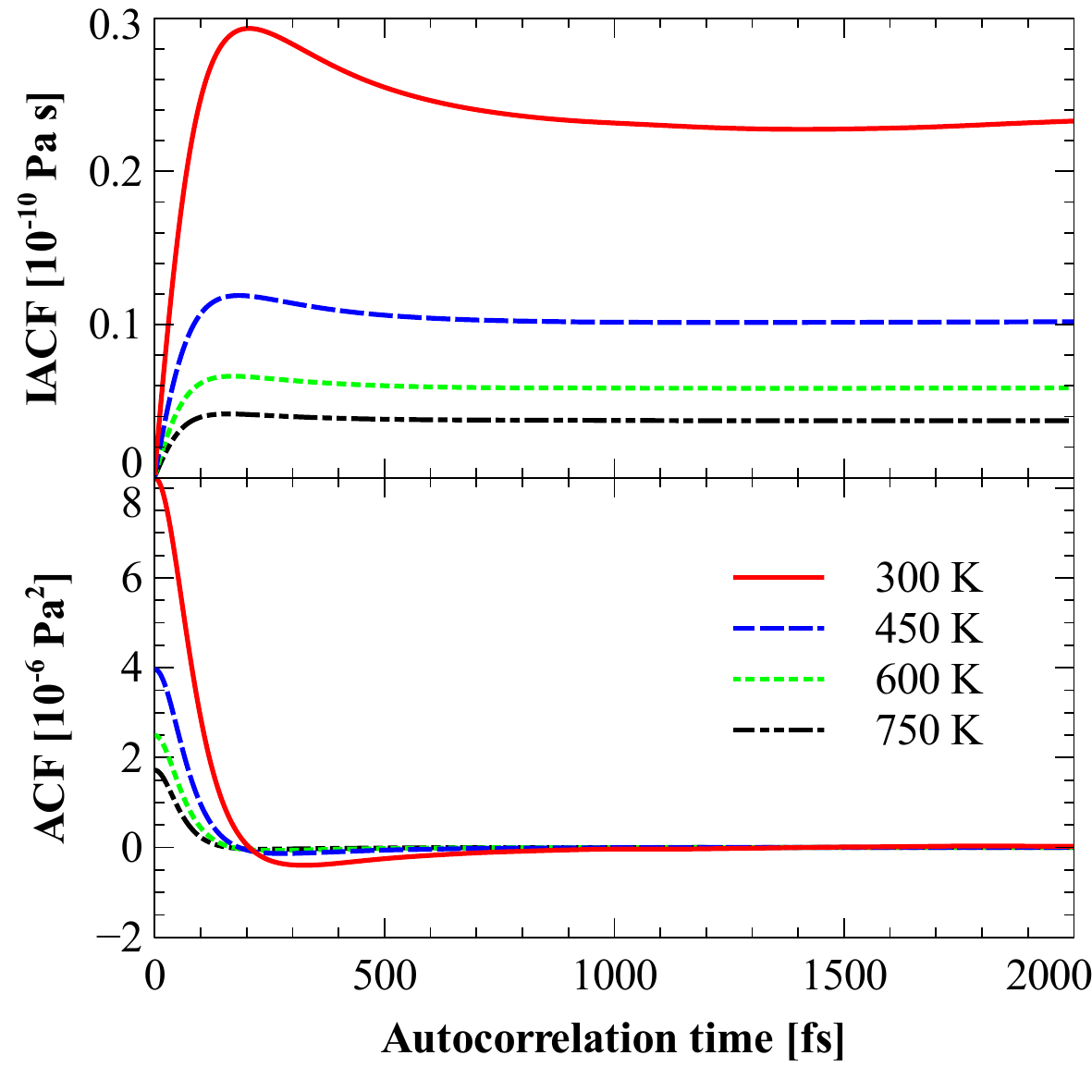}
	\label{fig:figure2a}
	\caption{0.5 bar}
\end{subfigure}
\begin{subfigure}{0.48\textwidth}
	\centering
	\includegraphics[width=0.9\linewidth]{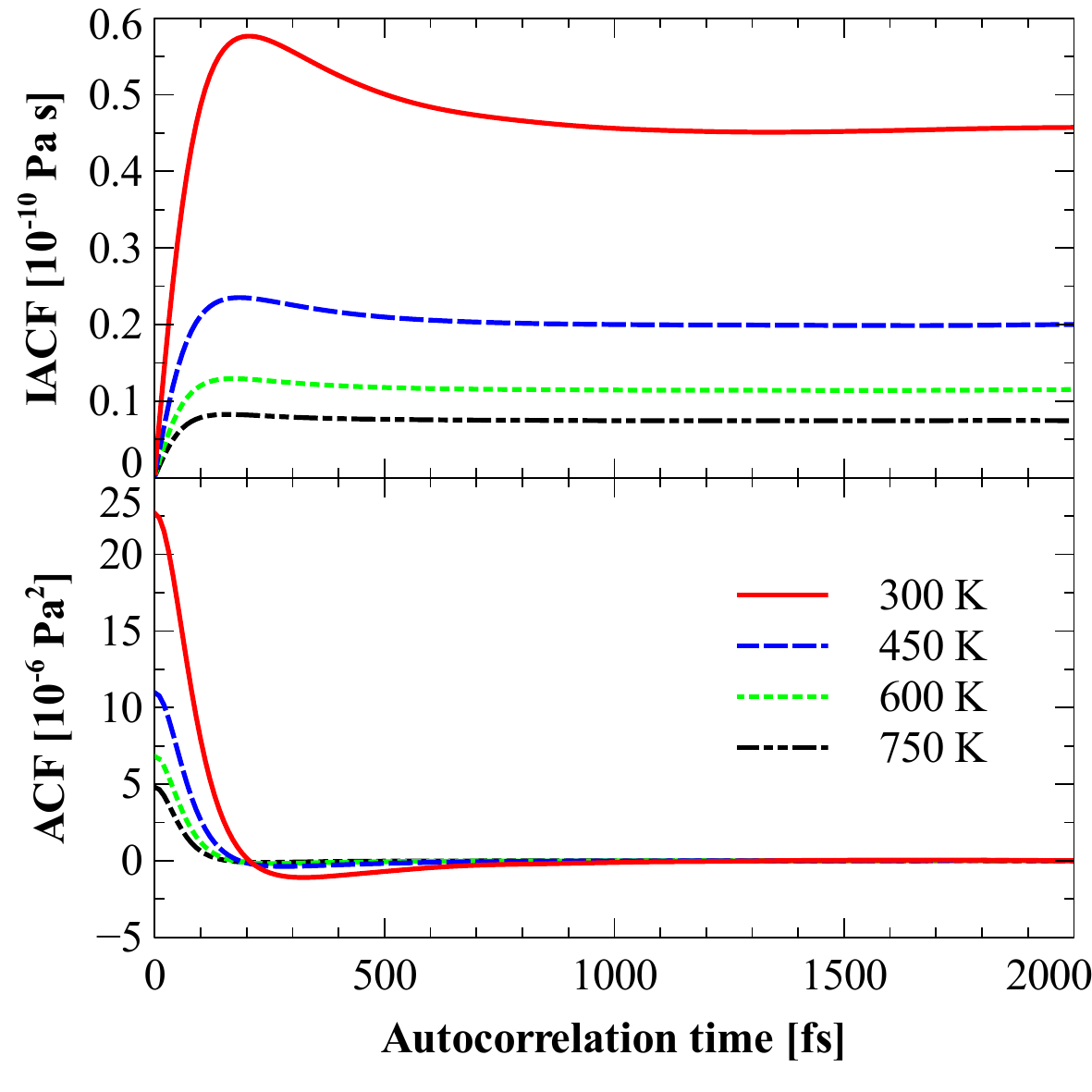}
	\label{fig:figure2b}
	\caption{0.7 bar}
\end{subfigure}

\begin{subfigure}{0.48\textwidth}
	\centering
	\includegraphics[width=0.9\linewidth]{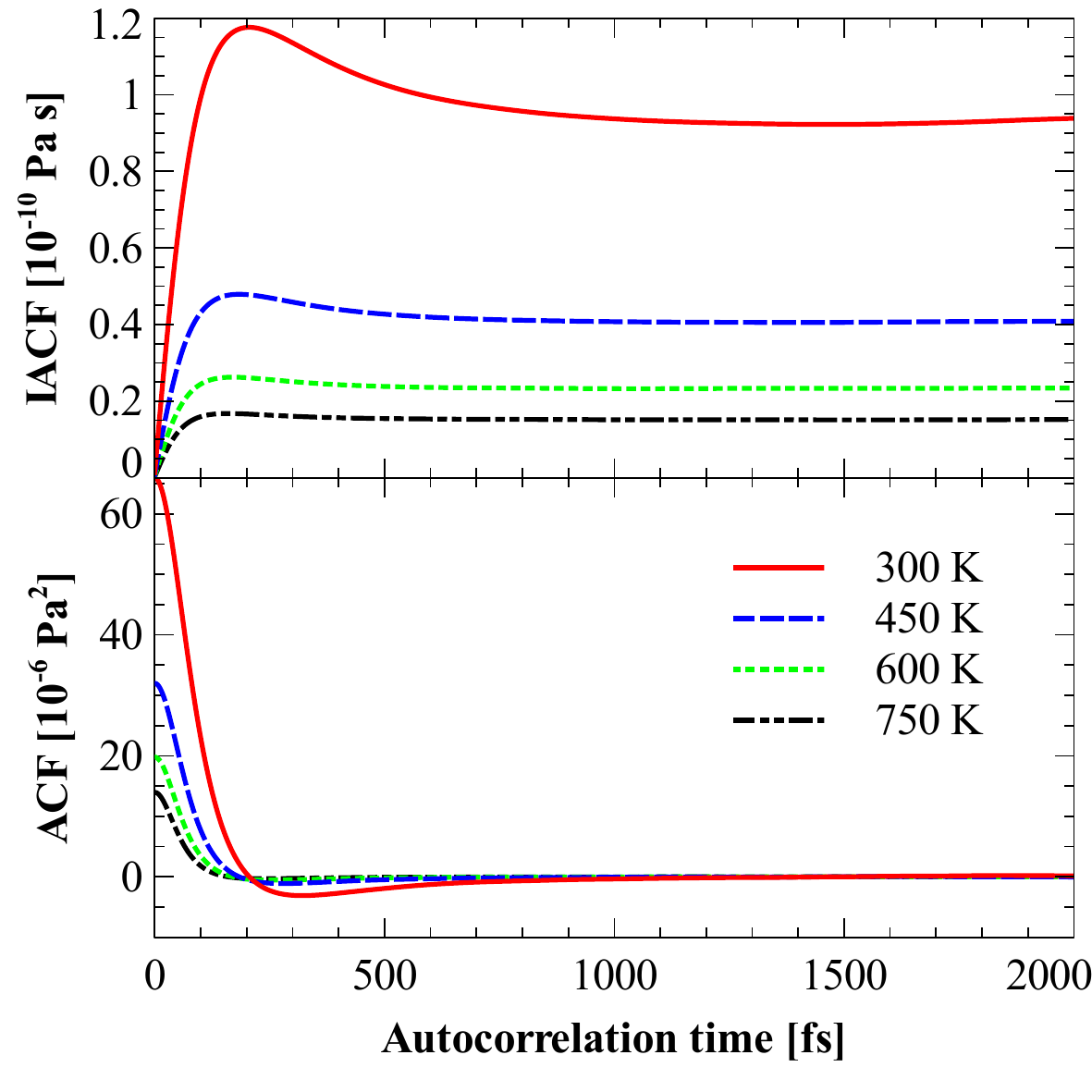}
	\caption{1.0 bar}
	\label{fig:figure2c}
\end{subfigure}
\begin{subfigure}{0.48\textwidth}
	\centering
	\includegraphics[width=0.9\linewidth]{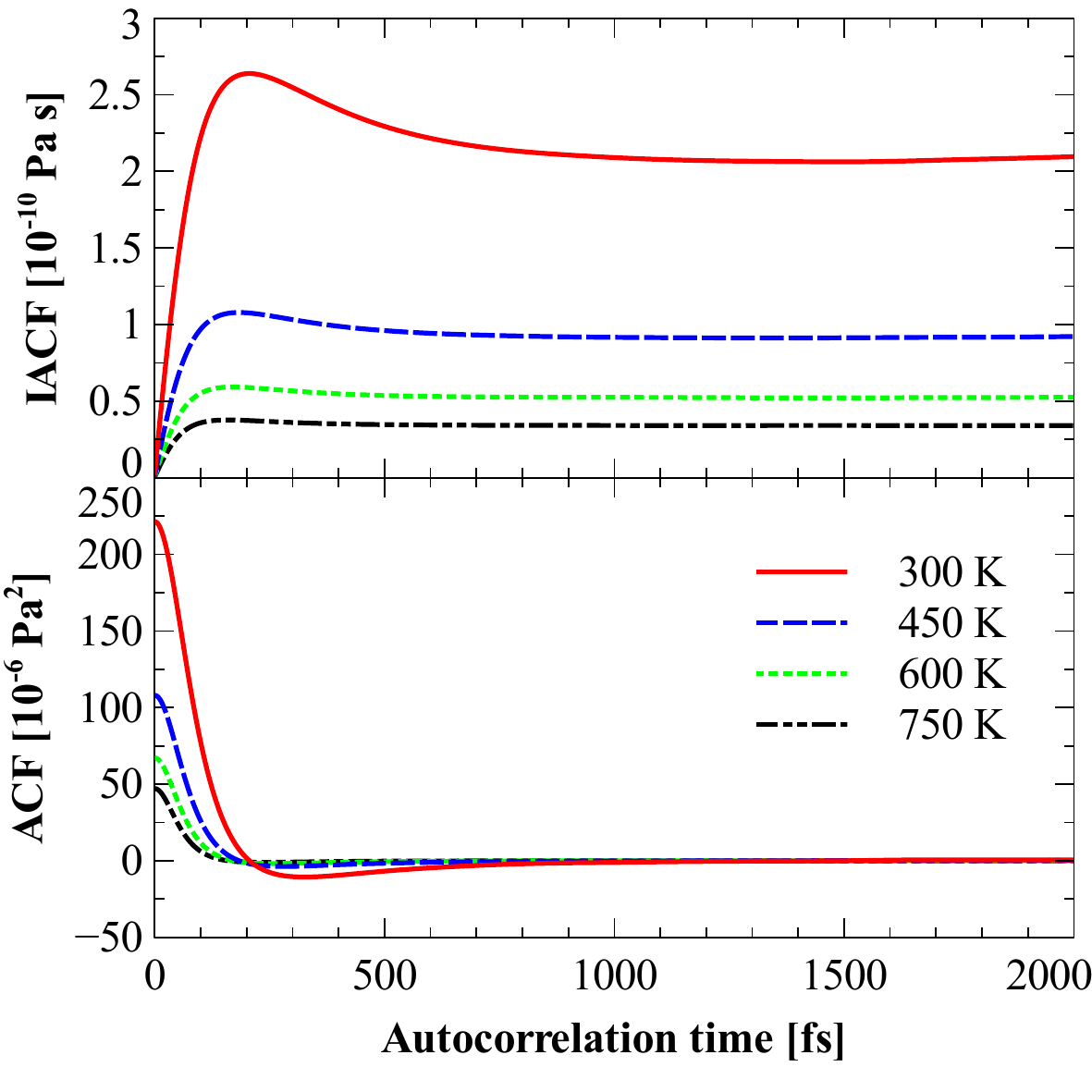}
	\caption{1.5 bar}
	\label{fig:figure2d}
\end{subfigure}

\caption{Plots of integral of autocorrelation function.} 
\label{fig:IACF}
\end{figure*}

Figure~\ref{fig:IACF} shows corresponding plots of autocorrelation function (ACF) and integral of autocorrelation function (IACF) with time. It can be observed that initially, the ACF decreases rapidly and shows a minimum. After that, it increases again and approaches the time axis. This first increasing and then decreasing behavior of ACF manifests a peak in corresponding IACF plots. This nature of ACF plots contrasts with that obtained at low temperatures for monatomic gases and that of diatomic gases at similar temperatures. In the former case, the plots show several oscillations caused by the formation of bound states \cite{meier2005transport, meier2004transport-1viscosity, meier2004transport-2selfdiffusion}. These oscillations decrease with the increase in temperature and disappear at sufficiently high temperatures. Therefore, these oscillations are not present in the present simulations. On the other hand, in the latter case, there are no oscillations at all as the bulk viscosity is caused by the energy exchange between translational and internal modes. 

\begin{figure}
\centering
\includegraphics[width=0.5\linewidth]{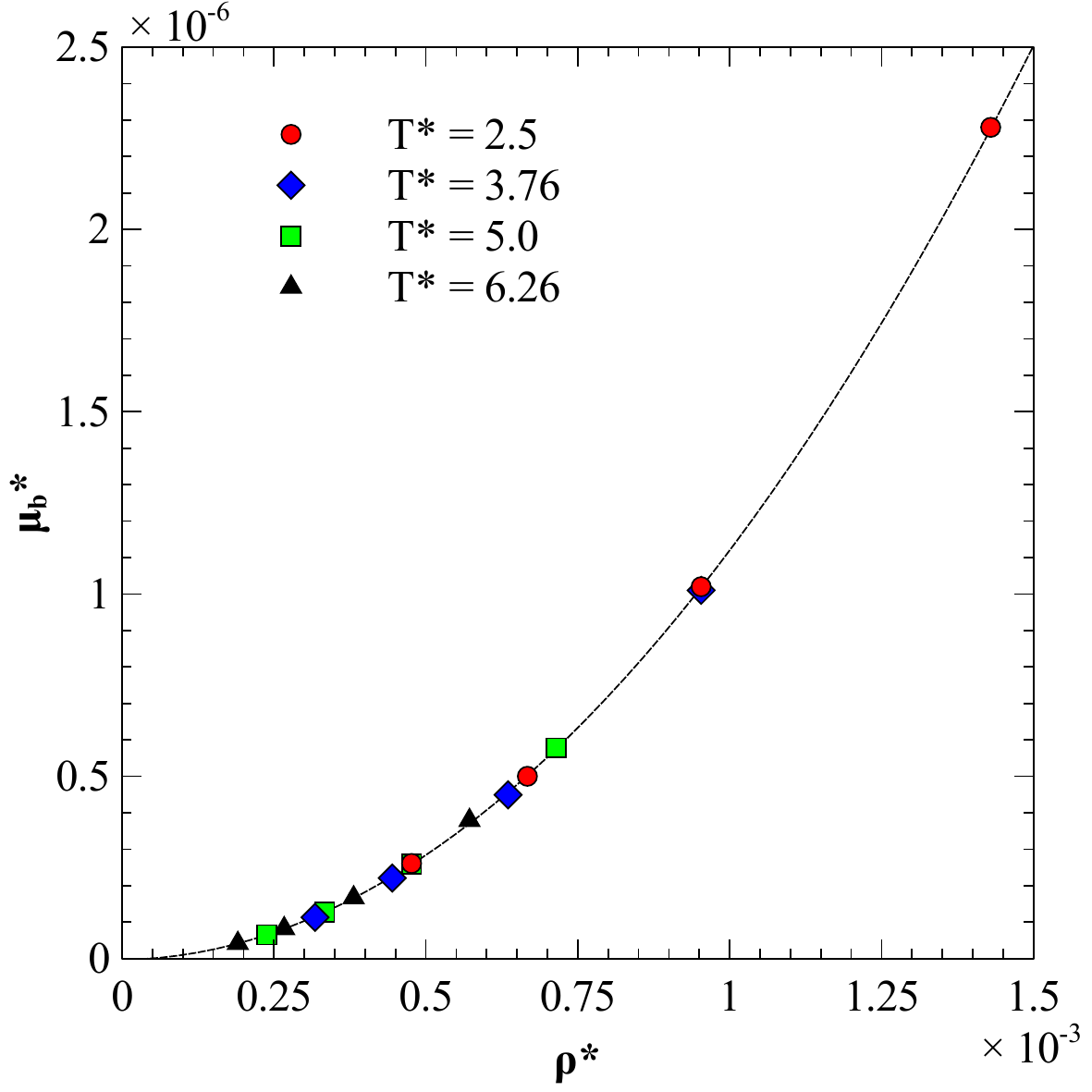}
\caption{Variation of $\mu_b^*$ with $\rho^*$. 
	The solid line represents the quadratic curve obtained by fitting data-points to the equation $y = ax^2+bx+c$. The details of curve fitting, viz., value of $a$, $b$, $c$, and goodness of fit $\chi^2$ are given in table~\ref{table:fitting parameters}. } 
\label{fig:reduced-units}
\end{figure}

\begin{figure}
\centering
\begin{subfigure}{0.48\textwidth}
\centering
\includegraphics[width=\linewidth]{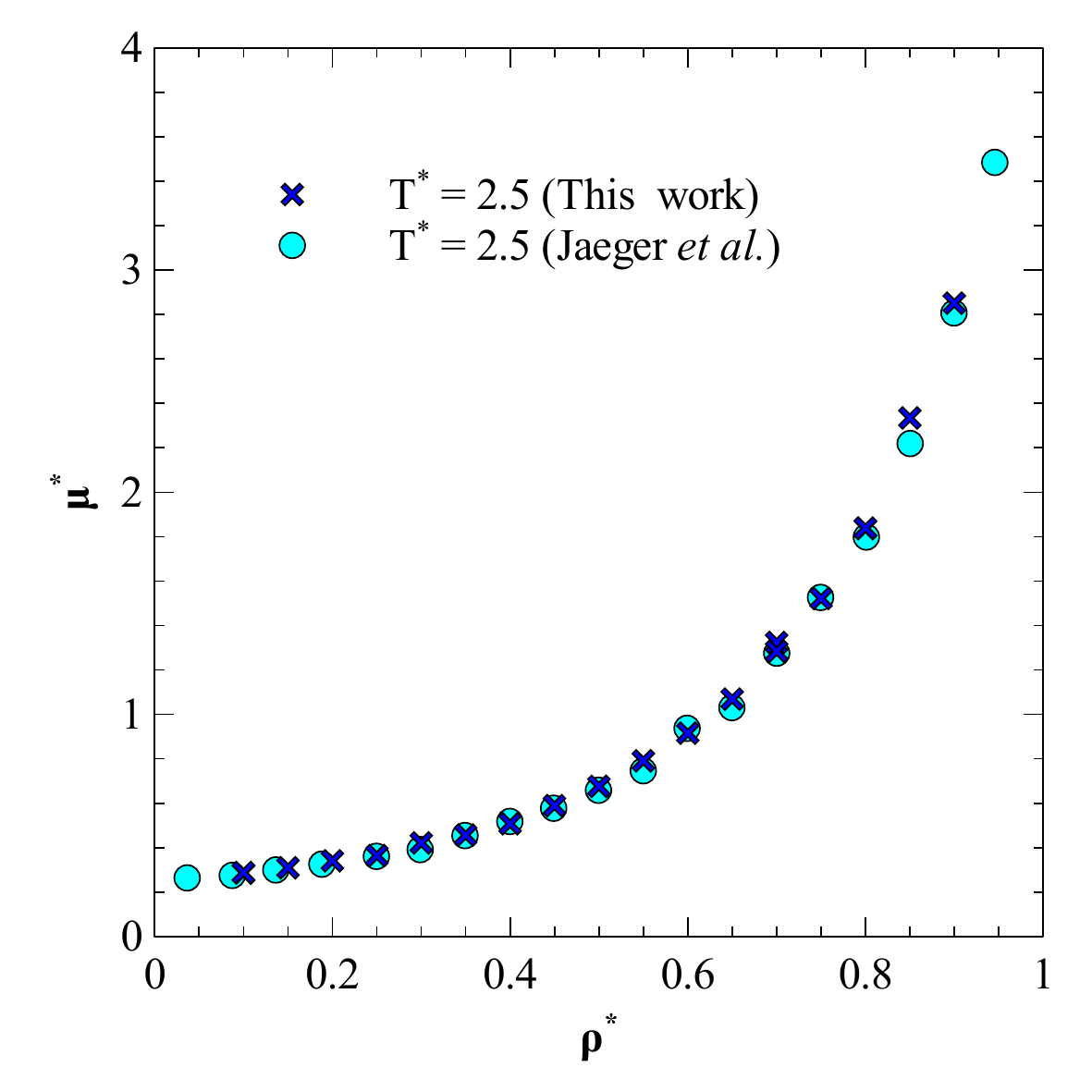}
\caption{}
\label{fig:dense-lj-mu-star}
\end{subfigure}
\begin{subfigure}{0.48\textwidth}
\centering
\includegraphics[width=\linewidth]{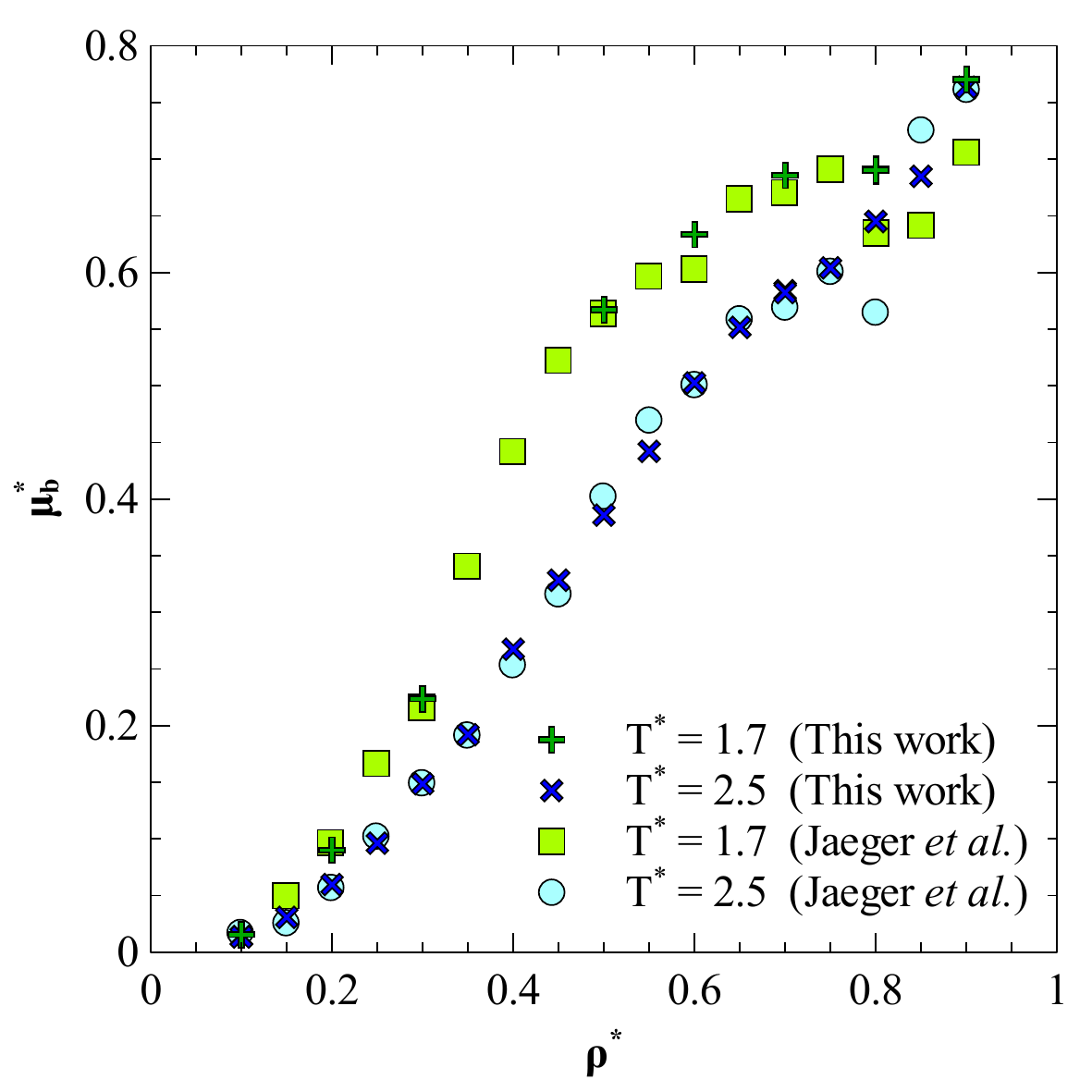}
\caption{}
\label{fig:dense-lj-mub-star}
\end{subfigure}
\caption{Shear and bulk viscosity of dense Lennard-Jones fluid.}
\label{fig:dense-LJ-results}
\end{figure}

To enable the extension of the obtained results for bulk viscosity of dilute argon gas to other noble gases, we have reduced the results in Lennard-Jones units \cite{plimpton2007lammps} 
The results are plotted in figure~\ref{fig:reduced-units}. It is interesting to note that the bulk viscosity becomes independent of temperature at dilute conditions as all the data points lie on the same quadratic curve. This nature of the curve is in contrast with that of dilute nitrogen gas at similar temperatures and pressures, for which bulk viscosity shows linear dependency on temperature.
This temperature-independent nature of monatomic gases at low densities is in line with the trends at higher densities (i.e., $\rho^*>0.01$, see figure~\ref{fig:dense-LJ-results})), where all the isotherms tend to converge to the same curve as the density decreases. Figure~\ref{fig:dense-LJ-results} shows shear and bulk viscosity of Lennard-Jones fluid calculated in a wide range of density.
However, the studies by Meier {\em et al.}\cite{meier2005transport} and Jaeger {\em at al.}\cite{jaeger2018bulk} were limited to $\rho^*>0.01$, and in this $\rho^*$ range, gases show density-dependent bulk viscosity values. To the best of our knowledge, no data have been reported so far for bulk viscosity of argon gas at conditions as dilute as $\rho*<0.001$. 


\begin{table}
	\begin{tabular}{|c|c|c|c|c|}
		\hline
		\textbf{Temperature (K)} & $\textbf{a}$ & $\textbf{b}$  & $\textbf{c}$ & \textbf{Goodness of fit} ($\chi^2$)  \\
		\hline\hline
		Fig. \ref{fig:variationwithpressureatalltemp}, 300 K &  0.9269069 &  -0.0252297 &  0.0166595   & 0.01833   \\
		\hline
		Fig. \ref{fig:variationwithpressureatalltemp}, 450 K & 0.4045920 &  0.0016059 &  3.1936196e-5 & 0.00016  \\
		\hline
		Fig. \ref{fig:variationwithpressureatalltemp}, 600 K & 0.2308508 &  0.0023903 & -0.0001509 & 0.01084  \\
		\hline
		Fig. \ref{fig:variationwithpressureatalltemp}, 750 K &  0.1517129 &  0.0002363 & -0.0003168 & 0.01268  \\
		\hline
		Fig. \ref{fig:reduced-units} &  1.0968105 &   2.6219489e-5 &  -3.6026319e-9 & 0.04878  \\
		\hline
	\end{tabular}
	\caption{Values of parameter $a$, $b$, $c$, and $\chi^2$ obtained form curve-fitting the data points of Fig.~\ref{fig:variationwithpressureatalltemp} and \ref{fig:reduced-units} in equation $y = ax^2+bx+c$. }
	\label{table:fitting parameters}
\end{table}	


At a first glance, the obtained nonzero bulk viscosity of monatomic gas may seem against the experimental results by Prangsma {\em et al.}\cite{prangsma1973ultrasonic}, who showed that bulk viscosity of dilute neon gas is zero. However, the uncertainty associated with their measurements was about 10\% or $O(10^{-6})$ Pa~s, which is significantly higher than the bulk viscosity of monatomic gases. Therefore, their results should not be interpreted with the impression that the bulk viscosity of dilute monatomic gases is exactly zero. 

The nonzero value of bulk viscosity of dilute argon gas can be explained as follows-- The origin of bulk viscosity in fluids is due to the fact that the pressure at non-equilibrium (i.e., mechanical pressure) is not the same as the pressure at equilibrium (i.e., thermodynamic pressure). There are primarily two mechanisms responsible for deviation of equilibrium pressure from non-equilibrium pressure--
(a) The slow exchange of translational energy with internal degrees of freedom. A brief review of this mechanism can be found somewhere else~\cite{sharma2019estimation}.
This mechanism is only present in polyatomic fluids as the monatomic molecules do not have rotational or vibrational energies.
(b) The second mechanism is the slow structural rearrangement of molecules after a change in the thermodynamic state of the fluid. This mechanism is primarily dominant in dense fluids. 
Here, the term `structural arrangement' refers to the configuration of a group of molecules packed together. 
When the density of the fluid is changed by either compression or expansion, the packing of molecules also needs to re-adjust accordingly. This process requires a finite time, and hence, becomes a source for bulk viscosity effects. 
Since the source of this mechanism of bulk viscosity is entirely due to the momentum transfer during the inter-molecular collisions \cite{madigosky1967density}, even monatomic fluids also have this mechanism. 
Nettleton \cite{nettleton1958intrinsic} referred to the first contribution as `apparent bulk viscosity' and second as `intrinsic bulk viscosity'. 
In the case of dilute noble gases, although the	apparent bulk viscosity is zero, the intrinsic bulk viscosity is finite. 

Figure~\ref{fig:variationwithpressureatalltemp} shows that the isotherms of bulk viscosity follow a quadratic trend. The isotherms are obtained by fitting bulk viscosity values in a quadratic curve, $ax^2+bx+c=0$. This trend is in agreement with the results of Gray and Rice~\cite{gray1964kinetic}, where they have applied the kinetic theory of simple liquids \cite{rice1961kinetic} and shown that the dense gas contribution or intrinsic bulk viscosity is proportional to the square of the density. Since for an ideal gas at constant temperatures, the density is directly proportional to the pressure of the gas, square dependency of intrinsic bulk viscosity also translates to the pressure. Noteworthy is the fact that only three out of the four points in each curve were used to calculate the parameters $a$, $b$, and $c$ of the quadratic equation. The excellent match of the fourth point with the value predicted by the quadratic curve shows that the variation of bulk viscosity with pressure is indeed of quadratic nature.   
\FloatBarrier
\section{Conclusions}

In this work, we performed precise Green-Kubo simulations to estimate bulk viscosity of dilute argon and dilute Lennard-Jones gas. The study revealed that bulk viscosity of dilute argon gas is $O(10^{-10})$ at atmospheric conditions, and the isotherms follow a quadratic trend in the temperature and pressure range considered here in this work. Sometimes, e.g., Ref.~\cite{pan2017role}, it is erroneously reported that bulk viscosity is exactly zero for dilute monatomic gases. Through the present study, we have demonstrated that the bulk viscosity of dilute monatomic argon gas is nonzero, though it is very small (roughly five orders of magnitude smaller than that of dilute nitrogen gas). The reason for the nonzero bulk viscosity of dilute monatomic gases is that the bulk viscosity of a fluid is manifested due to two mechanisms- slow relaxation of internal degrees of freedom and slow structural rearrangement of molecules. Though the former mechanism is absent in dilute monatomic fluids, the latter one is present and leads to nonzero bulk viscosity values. The absolute validity of Stokes' hypothesis holds good only in the ideal/hypothetical dilute gas limit. For all practical applications, where fluid can be safely considered as dilute, the assumption is true only in an approximate sense. 



\section*{Acknowledgments}

The authors gratefully acknowledge the use of computational resources provided by Dell HPC and AI Innovation Lab for carrying out computations required for this work. We acknowledge National Supercomputing Mission (NSM) for providing computing resources of 'PARAM Sanganak' at IIT Kanpur, which is implemented by C-DAC and supported by the Ministry of Electronics and Information Technology (MeitY) and Department of Science and Technology (DST), Government of India.

\section*{AUTHOR DECLARATIONS}

\subsection*{Conflict of Interest}
The authors have no conflicts of interest to disclose.

\subsection*{DATA AVAILABILITY}
The data that support the findings of this study are available from the corresponding author upon reasonable request.

\bibliography{Bibliography}  
\end{document}